\documentclass[conference]{IEEEtran}

\IEEEoverridecommandlockouts
\usepackage[utf8]{inputenc}
\usepackage{blindtext}
\usepackage{adjustbox}
\usepackage{graphicx}
\usepackage[hidelinks]{hyperref}
\usepackage{flushend}
\graphicspath{{figures/} {./}}

\usepackage{amsmath,amsfonts,amssymb}
\usepackage{caption}
\usepackage[labelformat=simple]{subcaption}

\DeclareCaptionLabelSeparator{periodspace}{.\quad}
\usepackage{lipsum}
\usepackage{placeins}

\usepackage[noadjust]{cite}
 \usepackage{tikz}

\usepackage[acronym]{glossaries}
\newacronym{gcd}{GCD}{Greatest Common Divisor}
\newacronym{lcs}{LCS}{Local Clock Set}
\newacronym{scc}{SCC}{Strongly Connected Components}
\newacronym{dag}{DAG}{Directed Acylic Graph}
\newacronym{csp}{CSP}{Communicating Sequential Processes}
\newacronym{eda}{EDA}{Electronic Design Automation}
\newacronym{sta}{STA}{Static Timing Analysis}

\definecolor{bluekeywords}{rgb}{0.13, 0.13, 1}
\definecolor{greencomments}{rgb}{0, 0.5, 0}
\definecolor{redstrings}{rgb}{0.9, 0, 0}
\definecolor{graynumbers}{rgb}{0.5, 0.5, 0.5}

\usepackage{lstautogobble}  
\usepackage{listings}
\begin{document}

\title{Yak: An Asynchronous Bundled Data Pipeline Description Language}

\author{
  \IEEEauthorblockN{Carsten Nielsen\IEEEauthorrefmark{1}}
  \IEEEauthorblockA{\textit{Institute of Neuroinformatics}\\
    \textit{University of Zurich and ETH Zurich}\\
    \textit{and} \\
    \textit{SynSense AG} \\
    Zurich, Switzerland\\
    carsten.nielsen@synsense.ai}
  \and
  \IEEEauthorblockN{Zhe Su}
  \IEEEauthorblockA{\textit{Institute of Neuroinformatics}\\
    \textit{University of Zurich and ETH Zurich}\\
    Zurich, Switzerland \\
    zhesu@ini.uzh.ch}
  \and
  \IEEEauthorblockN{Giacomo Indiveri}
  \IEEEauthorblockA{\textit{Institute of Neuroinformatics}\\
    \textit{University of Zurich and ETH Zurich}\\
    Zurich, Switzerland \\
    giacomo@ini.uzh.ch}

  \thanks{\IEEEauthorrefmark{1} Corresponding author.
    This work was partially supported by the European Research Council (ERC) under the European Union’s Horizon 2020 Research and Innovation Program Grant Agreement No. 724295 (NeuroAgents), and by the Electronic Component Systems for European Leadership (ECSEL) join undertaking Grant Agreement No. 876925 (ANDANTE).}
}

\maketitle


\begin{abstract}
  The design of asynchronous circuits typically requires a judicious definition of signals and modules, combined with a
  proper specification of their timing constraints, which can be a complex and error-prone process, using standard
  Hardware Description Languages (HDLs). In this paper we introduce \emph{Yak}, a new dataflow description language for
  asynchronous bundled data circuits. Yak allows designers to generate Verilog and timing constraints automatically,
  from a textual description of bundled data control flow structures and combinational logic blocks. The timing
  constraints are generated using the \gls{lcs} methodology and can be consumed by standard industry tools. Yak includes
  ergonomic language features such as structured bindings of channels undergoing fork and join operations, named value
  scope propagation along channels, and channel typing. Here we present Yak's language front-end and compare the
  automated synthesis and layout results of an example circuit with a manual constraint specification approach.
\end{abstract}


\bibliographystyle{IEEEtran}


\section{Introduction}
As the demand for energy-efficient and high-throughput hardware processing systems continues to grow in our current
computing landscape, the approach of asynchronous design is becoming more appealing, as it offers significant advantages
in terms of power consumption, noise emission, speed, and robustness compared to synchronous logic designs, as
demonstrated in numerous studies and
applications~\cite{Davies_etal14,Davies_etal18,Merolla_etal14a,Jiang_etal17,Chen_etal18a,Laurence12,Woods_etal97,Martin_etal97}.
Despite being a well-established discipline, the adoption of the asynchronous design approach is limited in both the
research and industrial landscape. Asynchronous design has been hindered by the lack of available \gls{eda} tools, and
by the knowledge and training required to efficiently build them. Despite these challenges, there is an ongoing effort
to address these issues and make asynchronous design more accessible with a fully open source flow from design
description to finished design~\cite{Ataei_etal21,Li_etal21a}.

Commercial \gls{eda} tools have been optimized for the synchronous logic design paradigm, which dominates today's
industrial scene. As such, they are not geared towards asynchronous design. For example, timing analysis of asynchronous
designs is not truly supported by standard tools, making it challenging for designers to create bundled data circuits
without advanced custom workflows or overly pessimistic delay buffer insertion on handshake request paths.

The asynchronous community has developed several tools to automate small and large parts of the design
flow~\cite{Ataei_etal21,Nowick_Singh15a,Nowick_Singh15}. However, most have been specific to the needs of a single lab
or commercial entity. Commercial entities typically choose not to open-source their tools which can make it a high-risk
decision to rely on as the company may choose to stop supporting it at any time with no other options available.

Other recent advances in timing analysis methods have also enabled the use of \gls{sta} tools to
analyze the timing of asynchronous pipelines. Using standard \gls{sta} allows using standard \gls{eda} 
tools with a normal flow and effectively outsources the development of hard components such as place and route
tools, power analysis etc. Making it more likely that it will be available for any chosen
process~\cite{Gimenez_etal18,Gimenez_etal19}.

Asynchronous design has been particularly popular in the neuromorphic community~\cite{Mead20}, with initial efforts
dating back to the 90s, to define and make use of the Address-Event Representation
(AER)~\cite{Lazzaro_etal93,Mahowald94b,Boahen98}, persevering to the more recent developments of large-scale and complex
spiking neural network processors such as SpiNNaker, TrueNorth, or
Loihi~\cite{Furber_Bogdan20,Merolla_etal14a,Davies_etal18}. However, the goal of neuromorphic designers is
typically to emulate the spike- (event-) based principles of neural processing using electronic circuits. As a
consequence, for the asynchronous design aspect, the focus is more on using circuits that serve the purpose of
processing event driven data in an asynchronous manner, than on improving their performance and efficiency. In this
respect, the lack of entry level \gls{eda} tools and the steep learning curve of asynchronous circuit design significantly
slows down asynchronous neuromorphic development, requiring years of study for new students to reach a productive level.

A large hurdle for new asynchronous circuit designers is the need to be deeply proficient in both gate level design and
timing analysis to even get started. To overcome this, a successful asynchronous design tool must require little to no
knowledge of timing analysis and allow describing circuits at an abstraction level at least as high as common HDL
processes, allowing designers to focus on the functionality of their circuits. This tool should be a minimal addon to
commercial tools, generating Verilog and timing constraints for use by commercial tools in the background. With back-end
tasks such as layout, routing, handling of PDKs being handled by a tool already available to most digital designers, the
development effort in the asynchronous community can be more narrowly focused on tasks and problems specific to
asynchronous circuit design. This minimizes the development burden on the asynchronous community and reduces potential bugs and
issues that may occur in the design process. Another factor that is important for promoting the adoption of an
asynchronous design tool by the community is its open-source nature: an open-source tool with a permissive license
would encourage wide adoption, including industry participation, and innovation. This encourages continued
development by multiple parties and minimizes the risk of obsoleteness caused by the original maintainers abandoning the
project. A permissive license allows companies to adopt the project for their own needs and contribute if they see
fit without any legal complications.

In addition to the above considerations, a successful asynchronous design tool should also have a user-friendly
interface that simplifies the design and debugging of novel circuits. This interface should allow for rapid prototyping
and experimentation, enabling designers to quickly test and refine their circuits. Good documentation and tutorials are
also necessary to attract designers to the community without requiring personal training. Distribution of the tool is
also critical, and distributing binary software that runs in many environments is a significant challenge.

Finally, having an open API in a scripting language for interacting with compiler internal data structures will allow
users to customize the tool to their needs without having to contribute to the core software. This approach promotes
customization and flexibility, making it an attractive tool for designers and researchers seeking to create complex
asynchronous pipelines. The result would be similar to what FIRRTL~\cite{Izraelevitz_etal17} is doing for synchronous
HDLs.

In this paper, we introduce \emph{Yak}, a new language for describing asynchronous pipelines that addresses many of the
challenges in asynchronous design. Yak provides a simple and intuitive syntax for specifying pipeline components and
their interactions, generating code in Verilog that can be used with industry-standard synthesis and place-and-route
tools in the back-end.

We demonstrate the effectiveness of Yak through a sample circuit utilizing the supported control elements, showing that
it can produce efficient and power-saving designs while reducing development time and complexity compared to other
asynchronous approaches. Yak is open source with a permissive license, and we provide comprehensive documentation
and tutorials to make it easy for designers to get started and contribute to the community.

\section{Yak Overview}
Yak is a dataflow language. Circuits are described as token flow graph where the nodes are flow control structures (FCS)
or combinational logic blocks and edges are channels connecting the FCS components. The designer specifies the token
flow graph directly and has full control over its structure. Yak does not synthesize the token flow graph from a
higher-level description, nor does it attempt to optimize the structure provided by the designer.

A key focus of Yak is to allow designers to specify the token control graph in the most convenient way possible. To this
end Yak attempts to provide a clear and intuitive syntax for specifying asynchronous channels carrying typed signals and
their connections to each other as well as to flow control structures. To reduce the amount of boilerplate code needed to
describe asynchronous pipelines, Yak includes type inference at the signal and channel level.

Yak also supports a range of advanced features, including support for hierarchical design through the ability to define
custom components. This allows designers to create modular, reusable designs and to tailor the language to
their specific needs.

Yak synthesizes to Verilog and generates SDC constraints, making it interoperable with the major \gls{eda} tools on the
market. As well as with open source tools such as Yosys and Icarus Verilog.

In this section we go over the most important pieces of syntax and features to give a feel for the language and allow
understanding of the example that will be presented later.
\subsection{Channels and Signal Types}
Channels are declared using the \texttt{chan} keyword and may be declared as part of a flow expression or stand on their
own. Channel declarations take an optional type specification. Channel types are aggregates of the signals carried on
the channels.

Signals are declared using the \texttt{sig} keyword and also take an optional type specification. If type specifications
are not provided, the Yak compiler will attempt to infer them.

Signal types are currently limited to unsigned logic of any width. Future releases of Yak are planned to include named
records with multiple signals.
\begin{lstlisting}[language={}]
// no type specifier
chan foo;
// bar must carry a signal named x of any type
chan bar : {sig x};
// baz must carry an 8 bit signal named x
chan baz : {sig x : logic[7:0]};
\end{lstlisting}
\begin{figure*}
  \centering
  \begin{subfigure}[b]{0.3\textwidth}
    \centering
    \includegraphics[width=\textwidth]{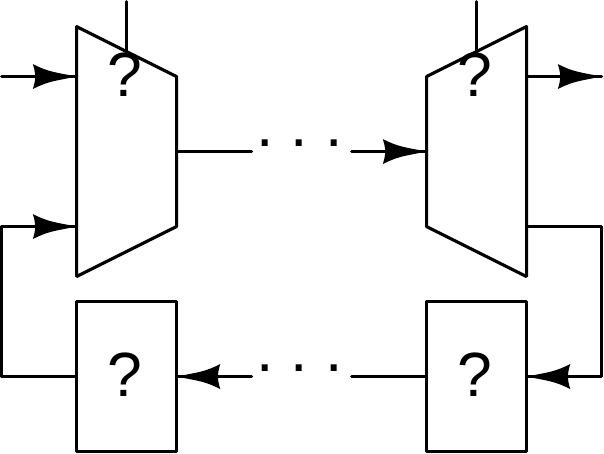}
    \caption{}%
    \label{fig:mux_ring}
  \end{subfigure}
  \hfill
  \begin{subfigure}[b]{0.3\textwidth}
    \centering
    \includegraphics[width=\textwidth]{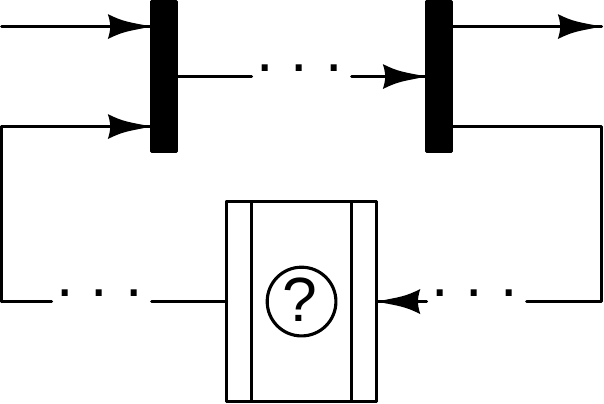}
    \caption{}%
    \label{fig:join_ring}
  \end{subfigure}
  \hfill
  \begin{subfigure}[b]{0.3\textwidth}
    \centering
    \includegraphics[width=\textwidth]{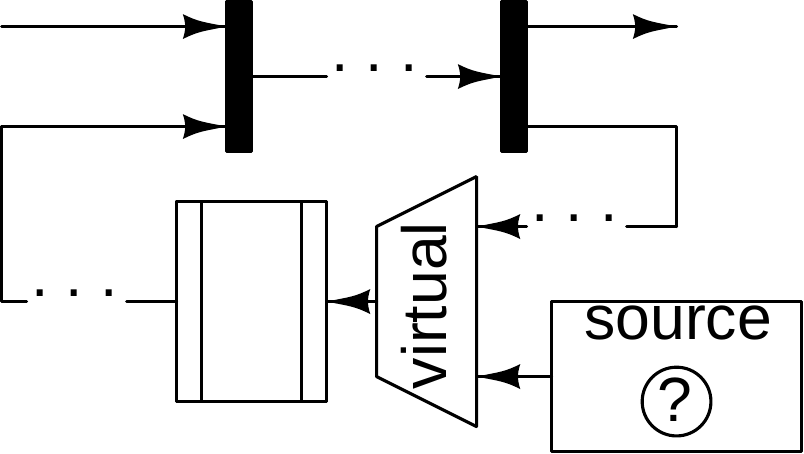}
    \caption{}%
    \label{fig:virtual_merge}
  \end{subfigure}
  \caption{Useful rings have a path for tokens to enter and exit the ring. A ring with a mux/merge entry has either a
    demux or fork exit (a) and a ring with a join must have a fork exit (b). Additionally a ring with only a join
    entry must have an initial token inside the ring after reset to avoid immediately deadlock on the join. In order
    to break all rings on the mux/merge entry to a ring where we can impose a type equality constraint we model a
    pipeline stage initialized with a token as an empty stage fed by a virtual merge component as depicted in (c).}%
  \label{fig:ring_fixing}
\end{figure*}
\subsection{Flow Controllers and Expressions}
Yak supports the following flow control structures which can be instantiated as built-in components.
\begin{itemize}
\item Join: Waits for a token to be available on all input channels, then produces one token on a single output channel.
\item Fork: Produces one token on every output channel for every token on its single input channel.
\item Merge: Produces one token on its single output channel for every token on either of its input channels.
  It is assumed that only one input channel will carry a valid token at any one time.
\item Mux: Waits for a token to be available on its input select channel and the channel indicated by the select token value, then
  produces one token on its single output channel.
\item Demux: Waits for a select token and a token on its single input channel, then produces a token on the output channel
  indicated by the select token value.
\item Arbit: Same as merge, without the channel exclusivity assumption.
\end{itemize}

Components and channels are connected using the \texttt{->} channel operator. This operator is the backbone of
describing circuits in Yak. Several syntactic sugars are provided to ease specification of graph structures using the
channel operator.

Channels may be connected to channels, for example \texttt{chan a -> chan b}. In this case we both declare the channels
and connected them at the same time. The expression is effectively a no-op, but is sometimes useful for readability.

The channel operator allows structured binding expressions on both sides. Every component in Yak expects some number of
channels as input and some number of channels as output. For example the join component expects two channels on its
input. Channels can be aggregated using square brackets into a structured binding. The join component must be
instantiated like so
\begin{lstlisting}[language={}]
[chan in0, chan in1] -> join() -> chan c;
\end{lstlisting}
And the fork component as the reverse
\begin{lstlisting}[language={}]
chan a -> fork() -> [chan b, chan c];
\end{lstlisting}

Channel aggregate expressions may themselves contain arbitrary channel flow expressions. For example
\begin{lstlisting}[language={}]
[[chan a, chan b] -> join(), chan c] -> join() -> chan d;
\end{lstlisting}
is a legal channel expression that aggregates channels a and b, joins them, and then aggregates the output of that join
with channel c to join them together into channel d.

Channels are implicitly created if multiple components are strung together using the \texttt{->} operator. For example
\begin{lstlisting}[language={}]
[chan a, chan b] -> join() -> fork() -> [chan c, chan d];
\end{lstlisting}
creates an implicit channel between the fork and the join components.

The \texttt{->} operator conveniently lets designers describe linear pipelines. For more complicated pipelines including
controlled components such as the \texttt{Mux} and \texttt{Demux} components Yak provides a way to specify side channels
like the select channel which is typically not part of the linear token flow.

\subsection{Components}
Components can be defined as
\begin{lstlisting}[language={}]
    name[inputs](side channels and parameters)[outputs] {
        // body
    }
\end{lstlisting}
Where inputs and outputs are lists of channels or channel aggregates separated by commas. Inputs and outputs declared here
become the input/output specification for the component when used in flow expressions. For example the component prototype for
the \texttt{join} component is
\begin{lstlisting}[language={}]
def join[chan a, chan b]()[chan c];
\end{lstlisting}

Inside the parentheses, it is possible to declare that the component requires static parameters or side channels. This is
used for the \texttt{Mux} and \texttt{Demux} components. The component prototypes for the mux and dmux are
\begin{lstlisting}[language={}]
def mux[chan a, chan b](chan s)[chan c];
def demux[chan a](chan s)[chan b, chan c];
\end{lstlisting}

When used in a flow expression we can use channels declared elsewhere as the channel carrying the select token.
\begin{lstlisting}[language={}]
[chan a, chan b] -> mux(chan s) -> chan c;
\end{lstlisting}
\subsection{Combinational Circuits}
Combinational blocks are instantiated using the \texttt{comb} keyword followed by statements enclosed in curly braces.
Inside the combinational blocks, a standard set of operators and conditional expressions are allowed. Temporary signals, also known
as variables, are declared using the \texttt{sig} keyword. Syntax is C-style.

Combinational circuits are transpiled directly into Verilog processes.

\begin{figure*}
  \centering
  \includegraphics[width=0.8\textwidth]{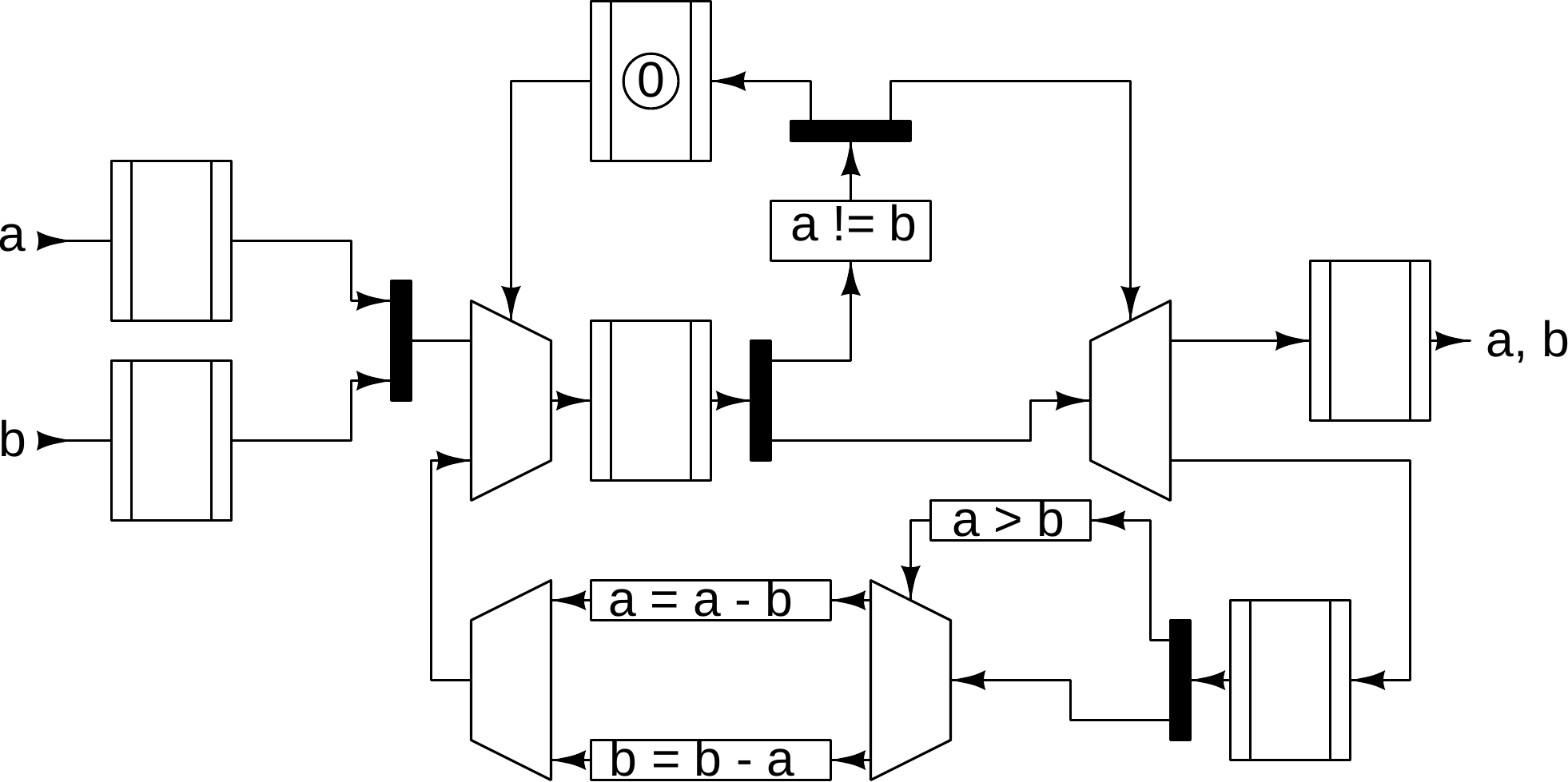}
  \caption{Asynchronous pipeline diagram of the \gls{gcd} circuit that is used in our example. The inputs \texttt{a} and \texttt{b}
    are first captured by initial pipeline stages and then joined to provide a token on which to compute the \gls{gcd}. The output
    is also capture in a pipeline stage in order to remove any effect of input and output constraint specification during
    physical implementation.}%
  \label{fig:gcd}
\end{figure*}

\subsection{Stateful Elements}
The builtin \texttt{reg()} component instantiates a pipeline stage with a stateful element that captures all the signals
present on the input channel when a token arrives. The component accepts an optional initialization parameter specifying
that a valid token is present on the output of this pipeline stage immediately after circuit reset. If the initializer
is specified, then all signals defined on the output channel are required to have an initial value specified. For
example
\begin{lstlisting}[language={}]
chan a -> reg(sig b : logic[7:0] = 0) -> chan c;
\end{lstlisting}
declares a pipeline segment where channel a is incident to a stateful element that initializes a signal b to 0.
The typing engine will check that a signal named b of the correct type is available on channel a.
\subsection{Inputs and Outputs}
When Yak components are transpiled to Verilog, the inputs and outputs of the generated Verilog module can be specified
with the builtin \texttt{input()} and \texttt{output} functions. The arguments to these functions are signal lists
specifying the signals that will become the inputs and outputs of the generated module. For example:
\begin{lstlisting}[language={}]
def foo[]()[] {
input(a, sig b : logic, sig c : logic[7:0]) ->
output(d, sig e : logic, sig f : logic[7:0]);
}
\end{lstlisting}
will be converted to the following Verilog module with predictable signal naming.
\begin{lstlisting}[language={}]
module foo (input req_a, output ack_a,
            input D_a_b, input [7:0] D_a_c,
            output req_d, input ack_d,
            output D_d_e, output [7:0] D_d_f)
begin
    assign req_d = req_a;
    assign ack_a = ack_d;
    assign D_d_e = D_a_b;
    assign D_d_f = D_a_c;
endmodule;
\end{lstlisting}
The typing engine will interpret the types specified on the input ports as fixed and unchangeable, and
verify that the types of the output signals are present on the channels connected to the outputs.

\subsection{Blackboxes}
Verilog modules may also be included for use in Yak source code as blackboxes. This is necessary for instantiating
memories and synchronous or asynchronous logic that does not need to be constrained using the SDC constraints generated
from Yak. When specifying a blackbox it is necessary to specify the type of every signal and channel input and output of
the blackbox. Blackboxes are instantiated using the built-in \texttt{blackbox(name, [inputs], [outputs])} function. For
example
\begin{lstlisting}[language={}]
blackbox(foo, chan a : {sig b : logic,
                        sig c : logic},
              chan d : {sig e : logic});
\end{lstlisting}
declares that the following Verilog module is to be instantiated in the transpiled design
\begin{lstlisting}[language={}]
module foo (input req_a, output ack_a,
            input D_a_b, input D_a_c,
            output req_d, input ack_d,
            output D_d_e)
\end{lstlisting}

Constraints for blackboxes must be specified manually by the designer. If the blackbox is an asynchronous bundled data
component, this can be done by either by specifying \gls{lcs} style constraints that bridge the gap between the last
pipeline stage defined in Yak and the first pipeline stage in the blackbox, or by any other constraint method that the
designer chooses. If the blackbox is a synchronous block that interfaces with the asynchronous Yak design the
constraints should be specified using either the LCS root clock from the last pipeline stage in the Yak design or an
actual real clock signal together with \verb|set_min/max_delay| constraints at the interface.
\subsection{Sources and Sinks}
It is sometimes necessary to instantiate components that unconditionally produce or consume tokens. This is done using
the built-in \texttt{source()} and \texttt{sink()} components. These take optional signal list declarations. If signals
are provided to the \texttt{sink()} component they must be available on the input channel of the instantiated sink components, while signals declared on the
\texttt{source()} component will be available on the output channel of the instantiated source component. For source components the available signals must be specified with a constant value.
\subsection{Scope Resolution and Typing}
One of the main drawbacks of specifying highly structural designs in classic HDLs such as Verilog and VHDL is the need
to specify the types of signals over and over. Especially in a pipelined design where one would have to specify the type
on every pipeline element as well as the wires connecting them. Yak solves this problem by automatically finding the
signals that must be present on any channel in the design as well as the type of that signal. For example in this design
\begin{lstlisting}[language={}]
chan foo -> comb {sig a = b + c;} -> bar ->
comb {e = a + b} -> baz;
\end{lstlisting}
scope resolution would find that there must be signals named \texttt{b} and \texttt{c} available on channel \texttt{foo},
and that signals \texttt{a} and \texttt{c} must be available on channel \texttt{bar}. All of the declared signals are available
for referencing on channel \texttt{baz}. If a signal is not used in later in the design, it will simply be dropped.

Several components place restrictions on the channel type inference. For example using the \texttt{reg()} component and
specifying an initial value will force the input and output signals of the register to carry exactly the signals
specified. When using \texttt{input()} and \texttt{source()} the designer effectively specifies that a component carrying
an exact set of signals on its output channel exists. The types on the output channels of these components are therefore
fixed and do not undergo inference. Similarly when using \texttt{output()} and \texttt{sink()} components the designer
specifies exactly the set of signals that should be available on the input channels of these components. Although in the
case of \texttt{sink()} it is allowed to omit the signal specification to obtain a universal sink that accepts any token
carrying any set of signals.
Using \texttt{blackbox()} creates a requirement constraint for all the specified signals on the
blackbox input channels and provides all the output signals of the blackbox on its output channels.

Yak uses live variable analysis, a standard procedure in all compilers, to work out the scope of signals by iteratively
propagating needed and provided signals until the token flow graph stabilizes. This removes the need to declare the same
signals over and over in different components, as one would be required to do if writing similar structural code in
Verilog.  After determining scope, type analysis requires a \gls{dag} of all the signal types and the operations that
are performed on them.

As almost all useful asynchronous circuits involve rings, building this graph necessitates breaking any rings by
removing some edges. However, if the wrong edges are chosen, it may be impossible to infer types for all edges in the
graph. To be able to perform type analysis we require that the outgoing edge of every root node in the graph is
specified by the designer. This happens naturally when using components such as \texttt{input()}, \texttt{source()}, and
\texttt{blackbox()}. If we find a loop and break it on a random edge in the loop, we may introduce a root node to the
graph which does not have types specified on its output edge. Yak avoids this problem by imposing an additional
constraint. Namely that the input channels to merge and mux components have identical types. With this constraint, we
can break rings whenever an input channel of a merge or mux component is part of the ring.

We observe that in order for a ring to perform useful computation, there must be a way for tokens to be introduced to,
and removed from the ring. The only components that can introduce tokens into a ring are joins, merges and muxes. In the
case where a join component introduces a token into a ring, there must also be an initial token in the ring in the form
of an initialized stateful element, or a merge, or mux which can introduce tokens into the ring formed by the join. If
none of these exist then the join component is guaranteed to deadlock as one of its inputs will never receive a
token. If an initial state token does exist in the ring, then this must have been specified using a \texttt{reg()}
component and the type is provided. In this case we transform the representation of the \texttt{reg()} node to a virtual
merge as depicted in Fig.~\ref{fig:virtual_merge}, allowing us to break all loops by considering only merges and muxes.

The Yak compiler's method for breaking these rings is to compute the \gls{scc} of the token flow graph. When we find an
\gls{scc}, we know that any edge crossing into the \gls{scc} must connect to a merge or mux. We then break the other,
internal, edge of this merge or mux to reduce the \gls{scc}. This procedure is repeated until the token flow graph is a
\gls{dag} and we can perform type analysis.
\begin{figure*}[h]
\begin{lstlisting}[language={}, caption={Structural description of the \gls{gcd} circuit in Yak.}, label={lst:gcd}, captionpos=b]

input(a, sig a : logic[7:0]) -> chan a;
input(b, sig b : logic[7:0]) -> chan b;

chan loop_s;

[[a -> reg(), b -> reg()] -> join(), chan loop_val] -> mux(loop_s) -> reg() -> fork() ->
[
        comb {sig s : logic = a != b;} -> fork() -> [reg(sig s : logic = 0) -> loop_s, chan output_s],
        demux(output_s) -> [reg() -> output(o, sig a : logic[7:0], sig b : logic[7:0]),
                            reg() -> fork() -> [comb {sig s : logic = a != b;} -> chan branch_s,
                                     demux(branch_s) -> [
                                                        comb {a = a - b;},
                                                        comb {b = b - a;}
                                                        ]
                                     -> merge() -> loop_val
                                     ]
                           ]
];

\end{lstlisting}
\end{figure*}

\subsection{Constraint Generation}
Generation of the \gls{lcs} SDCs for consumption by \gls{eda} tools requires that the constraints are generated in the
correct order. This is necessary because the \gls{lcs} constraints contain several generated clocks which must be
defined after the clock from which they are generated. To discover the correct order of constraint specification the Yak
compiler generates \gls{lcs} constraints by extracting all stage-to-stage paths in the design.  This is done for every
stage with the end result being a DAG for every stage. We call the stage for which the DAG is generated the
``root'' stage because this is where the \gls{lcs} root clocks will be defined for that stage. Note that the root stage
is not a root node in the DAG. It is the first node in the DAG to which every root has a path, and from which every
leaf can be reached.  In the DAG, the root nodes are other stages launching data that must be captured by the root
stage and the leaf nodes are stages that must capture data launched by the root stage. The compiler walks the DAG in
the forward direction from the root stage node to every leaf, generating all the generated clocks needed for setup
analysis in the correct order for every encountered token flow controller. And it then walks backward from the
root stage node to every root node and generates all the generated clocks needed for hold analysis.

In mux and demux components the data and request paths intersect. In addition to including them as nodes to be walked in
the DAGS capturing the stage-to-stage paths, we also generate stage-to-mux and stage-to-demux graphs where we treat the
mux or demux as the root stage node. These stage-to-mux and stage-to-demux graphs are processed in the same way as the
stage-to-stage graphs to generate the setup and hold constraints for the select data state elements in the mux and demux
components.

This process derives all the generated clocks in the timing constraints from the token flow graph described by the Yak
source. The constraints are emitted in the correct order that is required by the \gls{sta} tools without requiring
scripts that interact with an already elaborated design and inspect the tool design database.

\section{Example circuit}
In this section we compare asynchronous implementation methods of the \gls{gcd} circuit~\cite{Mardari_etal19}
implemented using the decoupled click element and its controllers. The \gls{gcd} circuit computes the greatest common divisor
of two input numbers using Euclid's algorithm. The asynchronous implementation is structured as a loop equivalent to
\begin{lstlisting}[language={}]
while (a != b) {
    if (a > b) {
        a = a - b;
    } else {
        b = b - a;
    }
}

output a,b;
\end{lstlisting}
We modified the \gls{gcd} circuit in~\cite{Mardari_etal19} slightly to showcase the full range of asynchronous flow control elements.
Firstly, we added a join to the input, which is not mandatory but allows us to demonstrate a circuit using a join.
Secondly, we placed pipeline registers on the inputs and outputs of the circuit, which enables a clearer understanding
of the \gls{lcs} constraints. These modifications also eliminate any impact of timing constraints on the input and output. A
diagram of the circuit is shown in Fig.~\ref{fig:gcd}.

The implementation of the circuit is written in Yak and is shown in listing~\ref{lst:gcd}. For the \gls{lcs} driven synthesis
the generated Verilog netlist and \gls{lcs} constraints are used to synthesize and perform place-and-route of the design using
Synopsis Design Compiler and Cadence Innovus.

We compare to a manual constraint specification flow which is similar with~\cite{Miorandi_etal17}. In this flow we use
the generic GTECH Synopsys library to implement the very low-level but technology-independent gate-level version of the
circuit.  During synthesis, we then only allow gate sizing and buffer insertion to be certain that no logic is changed
which would impact the timing paths.  The \verb |set_max_delay| command is then applied to all of the timing paths in
order to improve the result quality as meeting the coming \verb |set_min_delay| constraints would be easy for the tool
if the maximum delay is not constrained. The clock and reset paths are assigned higher weight of delay constraint during
technology mapping, which avoids minimum pulse width and hold time violations. After the first technology mapping, we
extract the delay of request path and data path between every pair of pipeline stages using the \verb|get_timing_paths|
and \verb|report_timing| commands. The \verb|set_min_delay| command is applied to delay a request path if it shorter
than the maximum delay of its corresponding data path after accounting for some margin.  This process is repeated until
all of bundling constraints are satisfied.

Three kinds of margins are used in \verb|set_min/max_delay| method, which means the delay of the request path is larger
than the data path by 5\%, 10\% and 20\% respectively. Usually when performing place and route with 5\% margin, the
tools cannot satisfy all of path constraints automatically. This then has to be solved by adding delay elements manually
which increases the design effort significantly. In practice, leaving 20\% margin is the common choice, especially in
applications which have higher requirements for robustness than performance, such as neuromorphic processors.

\subsection{Results}
Table~\ref{tbl:results} lists the post-route area and extracted timing simulation results of the \gls{gcd} circuit implemented
using \gls{lcs} constraints and the \verb|set_min/max_delay| method. For this example the two methods produce comparable
results. The difference being the higher design effort required when using the \verb|set_min/max_delay| method.
\begin{table}
  \centering
  \caption{Layout area and post-route extracted simulation results for the \gls{gcd} circuit using \gls{lcs} constraints generated
    by Yak and hand written \texttt{set\_min/max\_delay} constraints. Cycle time is defined as the time it takes a token to
    make a full loop through the \gls{gcd} circuit and latency is defined as the overhead incurred by the initial input
    stages, join component, and output pipeline stage. The margins listed are the margins applied to the request-to-data
    relative delay using the \texttt{set\_min/max\_delay} flow.}%
  \begin{tabular}{cccc}
    method & area (\(\mu\text{m}^2\)) & latency (ns)  & cycle time (ns) \\ \hline
    LCS & 320 & 0.53 & 1.01 \\
    5\% margin & 302 & 0.20 & 0.97 \\
    10\% margin & 311 & 0.51 & 1.06 \\
    20\% margin & 324 & 1.10 & 1.25 \\
  \end{tabular}
  \label{tbl:results}
\end{table}

\section{Conclusion}
This paper presents Yak, a dataflow language for describing asynchronous pipelines. The language includes features that
make specifying asynchronous circuits in a structural manner less tedious than using conventional HDLs, without relying
on synthesizing from a higher level description such as \gls{csp}. The language attempts to be a minimal addition to
flows using existing \gls{eda} tools, generating Verilog netlists and utilizing the \gls{lcs}
methodology~\cite{Gimenez_etal19} to generate timing constraints. When implemented using Cadence Innovus, these
automatically generated constraints produce equivalent results to manual buffer insertion, with significantly smaller
design effort. Yak and its compiler is open source with a nonrestrictive license and available at
\url{https://gitlab.com/neuroinf/yakl}. We plan to continue development of Yak to include a Python based API for
interacting with the token flow graph, enabling anyone to write custom graph walkers for performing optimizations, or
converting the controller graph to a format that can be consumed by other synthesis tools. The Python API will also
simulation at the token flow level, enabling test benches written in Python.

\bibliography{biblio/biblioncs}

\begin{thebibliography}{10}
\providecommand{\url}[1]{#1}
\csname url@samestyle\endcsname
\providecommand{\newblock}{\relax}
\providecommand{\bibinfo}[2]{#2}
\providecommand{\BIBentrySTDinterwordspacing}{\spaceskip=0pt\relax}
\providecommand{\BIBentryALTinterwordstretchfactor}{4}
\providecommand{\BIBentryALTinterwordspacing}{\spaceskip=\fontdimen2\font plus
\BIBentryALTinterwordstretchfactor\fontdimen3\font minus
  \fontdimen4\font\relax}
\providecommand{\BIBforeignlanguage}[2]{{%
\expandafter\ifx\csname l@#1\endcsname\relax
\typeout{** WARNING: IEEEtran.bst: No hyphenation pattern has been}%
\typeout{** loaded for the language `#1'. Using the pattern for}%
\typeout{** the default language instead.}%
\else
\language=\csname l@#1\endcsname
\fi
#2}}
\providecommand{\BIBdecl}{\relax}
\BIBdecl

\bibitem{Davies_etal14}
M.~Davies, A.~Lines, J.~Dama, A.~Gravel, R.~Southworth, G.~Dimou, and
  P.~Beerel, ``A 72-port 10g ethernet switch/router using
  quasi-delay-insensitive asynchronous design,'' in \emph{2014 20th IEEE
  International Symposium on Asynchronous Circuits and Systems}, 2014, pp.
  103--104.

\bibitem{Davies_etal18}
M.~Davies, N.~Srinivasa, T.-H. Lin, G.~Chinya, Y.~Cao, S.~H. Choday, G.~Dimou,
  P.~Joshi, N.~Imam, S.~Jain, Y.~Liao, C.-K. Lin, A.~Lines, R.~Liu,
  D.~Mathaikutty, S.~McCoy, A.~Paul, J.~Tse, G.~Venkataramanan, Y.-H. Weng,
  A.~Wild, Y.~Yang, and H.~Wang, ``Loihi: A neuromorphic manycore processor
  with on-chip learning,'' \emph{{IEEE} Micro}, vol.~38, no.~1, pp. 82--99,
  2018.

\bibitem{Merolla_etal14a}
P.~A. Merolla, J.~V. Arthur, R.~Alvarez-Icaza, A.~S. Cassidy, J.~Sawada,
  F.~Akopyan, B.~L. Jackson, N.~Imam, C.~Guo, Y.~Nakamura, B.~Brezzo, I.~Vo,
  S.~K. Esser, R.~Appuswamy, B.~Taba, A.~Amir, M.~D. Flickner, W.~P. Risk,
  R.~Manohar, and D.~S. Modha, ``A million spiking-neuron integrated circuit
  with a scalable communication network and interface,'' \emph{Science}, vol.
  345, no. 6197, pp. 668--673, Aug. 2014.

\bibitem{Jiang_etal17}
W.~Jiang, D.~Bertozzi, G.~Miorandi, S.~M. Nowick, W.~Burleson, and G.~Sadowski,
  ``An asynchronous noc router in a 14nm finfet library: Comparison to an
  industrial synchronous counterpart,'' in \emph{Design, Automation \& Test in
  Europe Conference \& Exhibition (DATE), 2017}, 2017, pp. 732--733.

\bibitem{Chen_etal18a}
W.~Chen, H.~Wu, S.~Wei, A.~He, and H.~Chen, ``An asynchronous energy-efficient
  cnn accelerator with reconfigurable architecture,'' in \emph{2018 IEEE Asian
  Solid-State Circuits Conference (A-SSCC)}, 2018, pp. 51--54.

\bibitem{Laurence12}
M.~Laurence, ``Introduction to octasic asynchronous processor technology,'' in
  \emph{2012 IEEE 18th International Symposium on Asynchronous Circuits and
  Systems}, 2012, pp. 113--117.

\bibitem{Woods_etal97}
J.~Woods, P.~Day, S.~Furber, J.~Garside, N.~Paver, and S.~Temple, ``Amulet1: an
  asynchronous arm microprocessor,'' \emph{IEEE Transactions on Computers},
  vol.~46, no.~4, pp. 385--398, 1997.

\bibitem{Martin_etal97}
A.~Martin, A.~Lines, R.~Manohar, M.~Nystrom, P.~Penzes, R.~Southworth,
  U.~Cummings, and T.~K. Lee, ``The design of an asynchronous mips r3000
  microprocessor,'' in \emph{Proceedings Seventeenth Conference on Advanced
  Research in VLSI}, 1997, pp. 164--181.

\bibitem{Ataei_etal21}
S.~Ataei, W.~Hua, Y.~Yang, R.~Manohar, Y.-S. Lu, J.~He, S.~Maleki, and
  K.~Pingali, ``An open-source eda flow for asynchronous logic,'' \emph{{IEEE}
  Design \& Test}, vol.~38, no.~2, pp. 27--37, 2021.

\bibitem{Li_etal21a}
R.~Li, L.~Berkley, Y.~Yang, and R.~Manohar, ``Fluid: An asynchronous high-level
  synthesis tool for complex program structures,'' in \emph{2021 27th IEEE
  International Symposium on Asynchronous Circuits and Systems (ASYNC)}, 2021,
  pp. 1--8.

\bibitem{Nowick_Singh15a}
S.~M. Nowick and M.~Singh, ``Asynchronous design--part 1: Overview and recent
  advances,'' \emph{{IEEE} Design \& Test}, vol.~32, no.~3, pp. 5--18, 2015.

\bibitem{Nowick_Singh15}
------, ``Asynchronous design--part 2: Systems and methodologies,''
  \emph{{IEEE} Design \& Test}, vol.~32, no.~3, pp. 19--28, 2015.

\bibitem{Gimenez_etal18}
G.~Gimenez, A.~Cherkaoui, G.~Cogniard, and L.~Fesquet, ``Static timing analysis
  of asynchronous bundled-data circuits,'' in \emph{2018 24th {IEEE}
  International Symposium on Asynchronous Circuits and Systems ({ASYNC})},
  2018, pp. 110--118.

\bibitem{Gimenez_etal19}
G.~Gimenez, J.~Simatic, and L.~Fesquet, ``From signal transition graphs to
  timing closure: Application to bundled-data circuits,'' in \emph{2019 25th
  {IEEE} International Symposium on Asynchronous Circuits and Systems
  ({ASYNC})}, 2019, pp. 86--95.

\bibitem{Mead20}
C.~Mead, ``How we created neuromorphic engineering,'' \emph{Nature
  Electronics}, vol.~3, no.~7, pp. 434--435, 2020.

\bibitem{Lazzaro_etal93}
J.~Lazzaro, J.~Wawrzynek, M.~Mahowald, M.~Sivilotti, and D.~Gillespie,
  ``Silicon auditory processors as computer peripherals,'' \emph{{IEEE}
  Transactions on Neural Networks}, vol.~4, pp. 523--528, 1993.

\bibitem{Mahowald94b}
M.~Mahowald, \emph{An analog {VLSI} system for stereoscopic vision}.\hskip 1em
  plus 0.5em minus 0.4em\relax Springer Science \& Business Media, 1994, vol.
  265.

\bibitem{Boahen98}
K.~Boahen, ``Communicating neuronal ensembles between neuromorphic chips,'' in
  \emph{Neuromorphic Systems Engineering}, T.~Lande, Ed.\hskip 1em plus 0.5em
  minus 0.4em\relax Norwell, MA: Kluwer Academic, 1998, pp. 229--259.

\bibitem{Furber_Bogdan20}
S.~Furber and P.~Bogdan, Eds., \emph{SpiNNaker: A Spiking Neural Network
  Architecture}.\hskip 1em plus 0.5em minus 0.4em\relax Boston-Delft: now
  publishers, 2020.

\bibitem{Izraelevitz_etal17}
A.~Izraelevitz, J.~Koenig, P.~Li, R.~Lin, A.~Wang, A.~Magyar, D.~Kim,
  C.~Schmidt, C.~Markley, J.~Lawson, and J.~Bachrach, ``Reusability is firrtl
  ground: Hardware construction languages, compiler frameworks, and
  transformations,'' in \emph{2017 {IEEE/ACM} International Conference on
  Computer-Aided Design ({ICCAD})}, 2017, pp. 209--216.

\bibitem{Mardari_etal19}
A.~Mardari, Z.~Jelčicová, and J.~Sparsø, ``Design and fpga-implementation of
  asynchronous circuits using two-phase handshaking,'' in \emph{2019 25th IEEE
  International Symposium on Asynchronous Circuits and Systems (ASYNC)}, 2019,
  pp. 9--18.

\bibitem{Miorandi_etal17}
G.~Miorandi, M.~Balboni, S.~M. Nowick, and D.~Bertozzi, ``Accurate assessment
  of bundled-data asynchronous nocs enabled by a predictable and efficient
  hierarchical synthesis flow,'' in \emph{2017 23rd {IEEE} International
  Symposium on Asynchronous Circuits and Systems ({ASYNC})}, 2017, pp. 10--17.

\end{thebibliography}

\end{document}